\documentstyle[sprocl]{article}
\def\Journal#1#2#3#4{{#1} {\bf #2}, #3 (#4)}

\def\PLB{{\em Phys. Lett.}  B}
\def\PRL{\em Phys. Rev. Lett.}
\def\PRD{{\em Phys. Rev.} D}

\def\be{\begin{equation}}
\def\ee{\end{equation}}
\def\bea{\begin{eqnarray}}
\def\eea{\end{eqnarray}}

\begin{document}
\title{(Vector) Glueballs and Charmonium Decay Revisited}
\author{George W. S. Hou}
\address{Department of Physics, National Taiwan University \\
       Taipei, Taiwan 10764, R.O.C. }
\maketitle\abstracts{
From the $0^{++}$ glueball candidate we argue that the
$1^{--}$ glueball $O$ to be near in mass to $J/\psi$.
This is supported by the recent null search for
$\psi^\prime \to \rho\pi,\ K^*\bar K$, while absence of distortion in
the energy scan of $J/\psi \to \rho\pi$ indicates that
$\Gamma_O \gg \Gamma_{J/\psi}$.
$O$ may also play a role in the OZI violating
$\bar p p \to \phi\gamma,\ \phi\pi$ production puzzle.
}

\section{Introduction: The Story of $O$}

We focus on the lightest vector glueball $O$,\cite{HS}
supposedly made up of 3 constituent gluons.
 It should mix less with $q\bar q$ mesons hence is cleaner than 2$g$ glueballs.

We are motivated by some recent developments:
i)  The $0^{++}$ glueball candidate\cite{AC}
$f_0(1500)$;
ii) Continued absence of $\psi^\prime\to \rho\pi$, $K^*\bar K$, with
similar situation emerging for VT modes;\cite{Gu}
iii) No distortion seen in energy scan of $J/\psi \to \rho\pi$.\cite{Gu,scan}
While i) and ii) support the existence of $O$,
iii) seems to cast doubt on 
the Brodsky-Lepage-Tuan (BLT) version\cite{BLT} of the vector glueball $O$.


%


First postulated by Freund and Nambu (FN)
for the OZI dynamics of $\phi\to \rho\pi$ decay,
$O$ is a ``closed string without quarks", hence a glueball.

In 1982, we were working on 3$g$ glueballs\cite{HS} in a 
potential model similar to the na\"\i ve quark model.
Taking the constituent gluon mass  to be
$m_g \sim 0.5$ GeV, 
we estimated that $m_O \cong 4.8 m_g \approx 2.4$ GeV.
To account for  $\Gamma(J/\psi \to \rho\pi) \simeq 1.1$ keV,
which disagreed with the prediction of FN,
we relaxed their $SU(4)$ assumption by allowing for
QCD motivated $q^2$ dependence, viz.
$f_{O\omega} : f_{O\phi} : f_{O\psi} = (\sqrt{2} : -1 : 1)f(q^2).
$ 
At that time MARK II had just reported for the first time the
``$\rho \pi,\ K^*\bar K$ anomaly".
That is,
although  $J/\psi \to \rho\pi$, $K^*\bar K$ decays
are quite sizable ($\sim 1\%$),
$\psi^\prime \to \rho\pi$, $K^*\bar K$ decays
were anomalously absent,
in contrast to the ``$15\%$ rule",
$
R_{\psi^\prime\psi} \equiv BR(\psi^\prime\to X)/BR(J/\psi\to X)
                     \sim 
                              0.15,
$ 
satisfied by most modes.
To explain this anomaly, we conveniently invoked $O$ pole dominance.\cite{HS}
Due to the proximity of $O$ to $J/\psi$ and its preference
to decay via $\rho\pi$,
it dominates over 3$g$ continuum for this mode,
but for most other modes the continuum is dominant.
$\psi^\prime$, however,  lies too far away from $O$ to be affected.
Thus, resonance enhancement of $J/\psi$ leads to the suppression
\be
{\Gamma(\psi^\prime\to \rho\pi) \over \Gamma(J/\psi\to \rho\pi)}
\cong {\left(m_{\psi}^2 - m_O^2 \over
m_{\psi^\prime}^2-m_O^2\right)^2} 
{f_{O\psi^\prime}^2\over f_{O\psi}^2}.
\ee
With $m_O \simeq 2.4$ GeV, the propagator ratio gives rise to a
factor of 4 suppression.

As the $\rho\pi$ anomaly deepened,
implying that $O$ must be rather close in mass
to $J/\psi$, 
BLT added the width factor and proposed that\cite{BLT}
\be
\vert m_O - m_\psi \vert < 80\ \mbox{MeV},\ 
\Gamma_O < 160\ \mbox{MeV},
\ee
was necessary to explain the anomaly.
Since $O$ is a 3$g$ state while $J/\psi$ a $c\bar c$ state,
this ``accidental degeneracy" may seem somewhat fortuitous.

\section{Crystal Barrel: $0^{++}$ Glueball Candidate and Its Impact}

An excess of $0^{++}$ states between 1350 -- 1750 GeV is observed
by the Crystal Barrel experiment and others at LEAR.
If we take $G \equiv f_0(1500)$ as mainly a $0^{++}$ glueball,\cite{AC}
we find that $m_{G} \simeq 1.4 - 1.5$ GeV,
which agrees well with lattice results.
Using potential model estimate $m_G \simeq 2.3 m_g$,
we infer the reasonable value
$ 
m_g \simeq 600 - 650\ \mbox{MeV} \simeq 2m_q,
$ 
implying that 
$m_O \cong 4.8 m_g \simeq 2.9$--$3.1\ \mbox{GeV},
$
right in the ballpark of Eq. 2!
The properties of $O$ have to be rechecked.


\vskip 0.25cm

Since $m_G + m_O > m_{\psi^\prime}$,
the decays\cite{HS} $J/\psi,\ \psi^\prime\to G+O\to \pi\pi+\rho\pi$ are forbidden.
But clearly $J/\psi \to \gamma G$ and $\psi^\prime \to \pi\pi + O$
search should continue.

Since the model demands that the $J/\psi \to \rho\pi$ width of 1.1 keV 
is saturated by the $J/\psi \to O\to \omega \to \rho\pi$ process,
one gets the relation,
\be
f(m_{J/\psi}^2) \cong 0.0247 \sqrt{m_{J/\psi}^2 - m_O^2}.
\ee
Taking for illustration $m_O \cong 2.9$ GeV, we find
the mixing angle
\be
\sin\theta_{\psi-O} \cong f(m_{J/\psi}^2)/(m_{J/\psi}^2 - m_O^2) \simeq 0.023,
\ee
which is reasonably small.
One then finds the widths for
$O\to \rho\pi,\ K^*\bar K$ to be roughly 2.2 MeV, 1.5 MeV,
while $O \to  e^+e^-,\ K\bar K,\ p\bar p$
are 2.7 eV (!), 0.08 MeV and 0.02 MeV, resp.
These numbers fit model prerequisites pretty well.
Assuming that BR$(O\to \rho\pi) \sim $ few--$10\%$,
we find that $\Gamma_O \simeq 30 - 100$ MeV,
a slightly narrow but still reasonable hadronic width.

\section{BES Results}

As reported at this meeting,\cite{Gu}  BES has
searched in vain for $\psi^\prime \to \rho\pi,\ K^*\bar K$ decay
with 3.5 $\times 10^{6}$ $\psi^\prime$.
Furthermore, a ``VT" anomaly seems to be emerging
(e.g. $R_{\psi^\prime\psi}(\omega f_2) < 0.035$).
These results seem to strengthen the case for $O$.
However, a recently published energy scan for $J/\psi \to \rho\pi$
over a 40 MeV interval sees no distortion from a nearby $O$ state.\cite{Gu,scan}
At first sight this may seem to rule out the BLT range of Eq. 2.
We wish to point out, however, that the null result should
probably be expected for $\Gamma_O \gg \Gamma_{J/\psi}$.
That is,
the two physical states are each admixtures of
$c\bar c$ and 3$g$ wavefunctions:
$\vert J/\psi \rangle = c_\theta \vert c\bar c\rangle + s_\theta \vert 3g\rangle,\  
\vert O \rangle = -s_\theta \vert c\bar c\rangle + c_\theta \vert 3g\rangle$.
Hence, the total cross sections for 
$e^+e^- \to \rho\pi$ under the $J/\psi$ and $O$ peaks should be equal,
and the $J/\psi$ peak stands out only because
it is extremely narrow.

In any case we hate to lose BLT hypothesis at this juncture
since it is now motivated by mass and width considerations from elsewhere.
To give up BLT, one would have to demand that
the mixing factor $f_{O\psi^\prime}^2/f_{O\psi}^2$
be suppressed.
This might in fact happen since both $O$ and $J/\psi$ are
ground states, while $\psi^\prime$ is a $2S$ state,
hence $O$ may overlap less well with $\psi^\prime$.

\section{Discussion}

We note that $J/\psi \to $ VP and VT and  $\eta_c \to $ VV modes
are quite prominent.
In our approach\cite{HS}
there are {\it two} vector states $1^{--}(0)$ and $1^{--}(2)$
and one $0^{-+}(1)$,
where number in parenthesis is the spin of
an octet 2$g$ pair. All three states turn out to be close to degenerate (!),
hence, resonance enhancement could occur for both $J/\psi$ and $\eta_c$.
The curiosity then is that, as the 3$g$ states decay, they seem to 
``remember" the spin configuration, i.e.
as the octet 2$g$ state converts to a $q\bar q$ pair,
it exchanges color and flavor with the other gluon via some
octet $q\bar q$ system.
Crude as it sounds, it may not be completely crazy.

With $m_O > 2.8$ GeV,
the $O\to \rho\pi$ width is no longer huge.
In a host of VP and VT decay modes,
$\rho\pi$, $\omega f_2$ etc. are the easiest to identify.

We wish to make a final remark on another anomaly 
coming from LEAR.
One normally expects 
$
R_X \equiv \sigma(\bar p p \to \phi + X)/\sigma(\bar p p \to \omega + X)
$ 
to be OZI suppressed, which is largely the case
{\it except} for $R_\pi,\ R_\gamma \simeq 0.1,\ 0.24$.
The former process occurs via a $^3S_1$ state,
while the latter is from a $^1S_0$ state.
In these rather controlled processes,
$O$ dominance ($O$-$\phi$ and $O$-$\omega$ mixing)
would imply the simple ratio $R_X \simeq 1/2$.
The likely presence of more channels for $\omega + X$
would drive this number down,
in accord with observation.


\section*{References}
\begin{thebibliography}{99} 
\bibitem{HS}W.S. Hou and A. Soni, \Journal{\PRL}{50}{569}{1983};
    \Journal{\PRD}{29}{101}{1984}.
\bibitem{AC}C. Amsler and F.E. Close, \Journal{\PLB}{353}{385}{1995}.
\bibitem{Gu}Talks by Y.F. Gu (BES Collab.) , this proceedings.
\bibitem{scan}J.Z. Bai {\it et al} (BES Collab.), \Journal{\PRD}{54}{1221}{1996}.
\bibitem{BLT}S.J. Brodsky, G.P. Lepage, S.F. Tuan, \Journal{\PRL}{59}{621}{1987}.
\end{thebibliography}
\end{document}